\documentclass[a4paper]{jpconf}
\usepackage{graphicx}
\begin{document}

\title{Excited Fermions at H1}

\author{E. Sauvan\footnote{On behalf of the H1 Collaboration}}

\address{CPPM, IN2P3-CNRS et Universit\'e de la M\'editerran\'ee, 163 Av. de Luminy\\
 F-13288 Marseille, France}

\ead{sauvan@cppm.in2p3.fr}

\begin{abstract}
We present a search for excited neutrinos and electrons  using all data collected by the H1 experiment at HERA at a center-of-mass energy of $320$~GeV with an integrated luminosity of up to $435$~pb$^{-1}$. No evidence for excited neutrino or electron production is found. Mass dependent exclusion limits are determined for the ratio of the coupling to the compositeness scale, $f/{\Lambda}$. 
These limits greatly extend the excluded region to higher masses than has been possible in previous searches.

\end{abstract}

\section{Introduction}

The existence of three distinct generations of fermions and the hierarchy of their masses motivates the possibility of a new scale of matter yet unobserved.
An unambiguous signature for a new scale of matter would be the direct observation of excited states of fermions ($f^*$), via their decay into a gauge boson and a fermion. Effective models describing the interaction of excited fermions with standard matter have been proposed~\cite{Hagiwara:1985wt,Boudjema:1992em,Baur:1989kv}.
In the models~\cite{Hagiwara:1985wt,Boudjema:1992em} the interaction of an $f^*$ with a gauge boson is described by a magnetic coupling proportional to $1/\Lambda$ where $\Lambda$ is a new scale. Proportionality constants $f$, $f'$ and $f_s$ result in different couplings to $U(1)$, $SU(2)$ and $SU(3)$ gauge bosons.
Electron (or positron) and proton collisions at very high energies provide an excellent environment to look for excited states of first generation fermions. 
In particular, these excited neutrinos ($\nu^*$) and electrons ($e^*$) could be singly produced through $t$-channel $W$ or $\gamma$/$Z$ boson exchange, respectively.

In this paper we present a search for both excited neutrinos and electrons using all HERA collider data of the H1 experiment. 
The data collected at electron and proton beam energies of $27.6$~GeV and $920$~GeV respectively corresponds to an integrated luminosity of up to $435$~pb$^{-1}$. 
The excited neutrinos and electrons are searched for through all their electroweak decays into a fermion and a gauge boson ($\gamma$, $W$ and $Z$). 

\section{Search for excited neutrinos}

In this search, hadronic and leptonic decay channels of the $W$ and $Z$ bosons are considered (see table~\ref{tab:exc_fermions}). 
The analysis covers $\sim$ $90$\% of the total branching ratio of $\nu^*$ decay. 
The selection criteria of the three main decay channels are described in the following.

In the $\nu\gamma$ resonance search, the principal signature is an isolated electromagnetic cluster in events with missing tranverse momentum ($P_{T}^{miss}$). Backgrounds arise from charged current (CC) events with an isolated ${\pi}^0$ or initial or final state radiation.
The presence of a neutrino is inferred by requiring $P_{T}^{miss} > 15$~GeV.
A photon is tagged by identifying an electromagnetic cluster with $P_T^\gamma >$ $20$~GeV. No well measured track should point to the electromagnetic cluster.
Radiative CC events are suppressed by requiring that the four-momentum transfer squared determined from the electromagnetic cluster be $\log (Q^2_\gamma) > 3.5$ GeV$^2$.

\begin{table}[htbp]
\caption{\label{tab:exc_fermions}Observed and predicted event yields for the studied decay channels.
  The errors on the prediction include model uncertainties and experimental systematic errors added in quadrature.}

\begin{center}
  \vspace*{5pt}
\begin{tabular}{l c c c }
\hline
Selection & Data & SM & Efficiency $\times$ BR\\
\hline
${\nu}^{*} {\rightarrow} {\nu}{\gamma}$ & $9$ & $15~{\pm}~4$ & $50$ \%\\
\hline
${\nu}^{*} {\rightarrow} {e}{W_{{\hookrightarrow}q\bar{q}}}$ & $198$ & $189~{\pm}~33$ &  $30$--$40$ \%\\
\hline
${\nu}^{*} {\rightarrow} {\nu}{Z_{{\hookrightarrow}q\bar{q}}}$ & $111$ & $102~{\pm}~24$ &  $40$ \%\\
\hline
${\nu}^{*} {\rightarrow} {e}{W_{{\hookrightarrow}\nu\mu}}$  & $0$ & $0.54~{\pm}~0.04$& $3$--$4.5$ \%\\
\hline
${\nu}^{*} {\rightarrow} {e}{W_{{\hookrightarrow}{\nu}e}}$ & $0$ & $0.6~{\pm}~0.3$ & $4$--$6$ \% \\
\hline
 ${\nu}^{*} {\rightarrow} {\nu}{Z_{{\hookrightarrow}ee}}$ & $0$ & $0.12~{\pm}~0.04$ &  $2$ \%\\
\hline
\multicolumn{4}{c}{\vspace{-0.2cm}}\\
\hline
${e}^{*} {\rightarrow} {\nu}{W_{{\hookrightarrow}q\bar{q}}}$ & $172$ & $175~{\pm}~39$ & $\sim 40$ \% \\
\hline
${e}^{*} {\rightarrow} {e}{Z_{{\hookrightarrow}q\bar{q}}}$ & $351$ & $318~{\pm}~64$ & $\sim 45$ \%\\
\hline                                        
${e}^{*} {\rightarrow} {e}{\gamma}$ & $112$ & $125~{\pm}~19$ & $60$--$70$ \%\\ 
\hline
\end{tabular}
\end{center}
\end{table}

For the $\nu Z_{{\hookrightarrow}q\bar{q}}$ resonance  search, the dominant SM background consists of multi-jet CC events with a moderate contribution from photoproduction.
The presence of a neutrino is inferred from substantial missing momentum $P_{T}^{miss} > 20$~GeV. 
We use a subsample of events with at least two jets, each having a transverse momentum larger than $20$ and $15$~GeV, respectively, and a polar angle between $5^{\circ}$ and $130^{\circ}$. 
At low $P_{T}^{miss}$, cuts on the variable  $\sum_i E^i - P^i_z$, where the sum runs over all visible particles, and on the ratio $V_{ap}/V_{p}$ of transverse energy flow anti-parallel and parallel to the hadronic final state~\cite{Adloff:2003uh} are used to suppress CC and photoproduction events. 
The $Z$ candidate is reconstructed from the combination of two jets with an invariant mass closest to the nominal $Z$ boson mass.
 
\begin{figure}[htbp] 
  \begin{center}
\includegraphics[width=.36\textwidth]{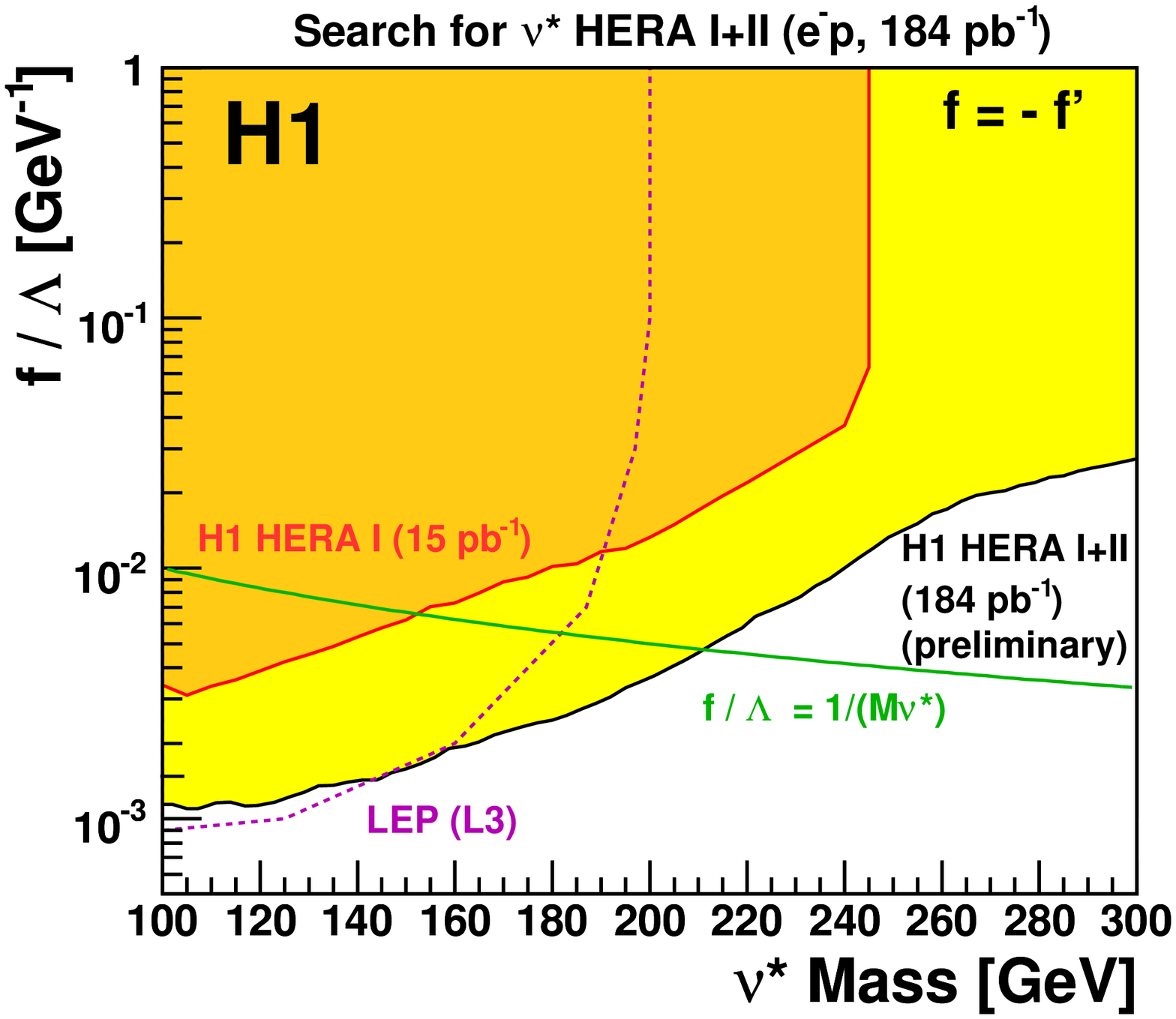}\put(-25,30){{\bf (a)}}
\includegraphics[width=.36\textwidth]{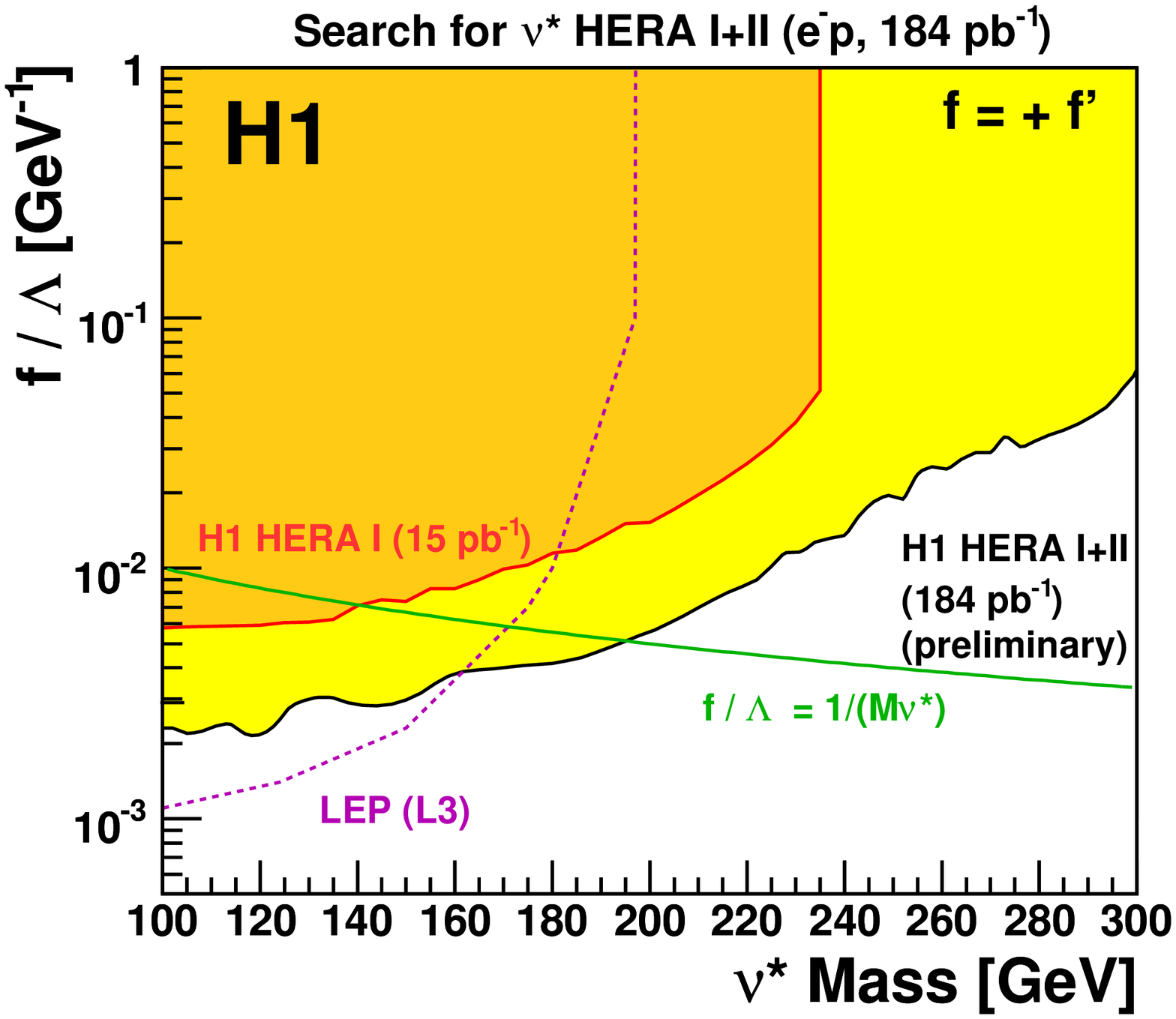}\put(-25,30){{\bf (b)}}
  \end{center}
  \vspace*{-17pt}
  \caption{Exclusion limits on the coupling $f/\Lambda$ at $95$\% C.L. as a function of the mass of the excited neutrino with the assumptions $f = -f'$ (a) and $f = +f'$ (b), respectively. 
}
\label{fig:nustar}  
\end{figure}

In the $eW_{{\hookrightarrow}q\bar{q}}$ channel, multi-jet neutral current (NC) events constitute the main contribution from SM processes.
Events are selected by the presence of an electron in the LAr with  $Q^2_e > 2500$~GeV$^2$ or a tranverse momentum $P_T^e$ greater than $25$~GeV. The restriction on the electron polar angle to the forward region of the LAr ($5^\circ< \theta^e < 90^\circ$) removes a large part of the NC DIS contribution.
The presence of at least two high $P_T$ jets with $P_{T}^{jet1,\; jet2} > 20, 15$~GeV and  $5^{\circ} < \theta^{jet1,\; jet2} < 130^{\circ}$ is required.
A $W$ candidate is then reconstructed in those events from the combinaison of two jets with an invariant mass closest to the nominal $W$ boson mass.

The event yields observed in each decay channel are summarised in table~\ref{tab:exc_fermions}. The observed event yields are in good agreement with SM expectations. 
No deviation is also observed in invariant mass distributions of excited neutrino candidates reconstructed in each decay channel.
Since there is no evidence for excited neutrino production, new bounds on the $\nu^*$  mass as a function of $f/\Lambda$ have been derived. 
They are presented in figure~\ref{fig:nustar}(a) and (b), under the two conventional assumptions $f = - f'$ and $f = + f'$, respectively. 
Assuming $f/\Lambda = 1/M_{\nu^*}$ and $f = - f'$, $\nu^*$ masses below $211$~GeV are ruled out. 

This result is presently the most stringent world limit for $\nu^*$ production and demonstrates the unique HERA sensitivity for $\nu^*$ masses beyond LEP reach.

\section{Search for excited electrons}

In the search for $e^*$, $W$ and $Z$ bosons are reconstructed only in the hadronic decay channel.
The total branching ratio accessed is $\sim$ $80$\%.

The signature in the $e\gamma$ channel consists of two isolated, high transverse momentum electromagnetic clusters. The principal SM background arise from elastic and inelastic Compton events.
Two isolated electromagnetic clusters in the  polar angle $5^\circ < \theta_{e1,e2} < 130^\circ$ are required, with a transverse momentum larger than $20$ and $15$~GeV, respectively.
To further suppress contributions from Compton events, the sum of the energies of the two electromagnetic clusters has to be greater than $100$~GeV and the sum of their transverse momentum should be larger than $75$~GeV. 

The selection criteria used in the $\nu W_{{\hookrightarrow}q\bar{q}}$ and $eZ_{{\hookrightarrow}q\bar{q}}$ channels are similar to those of the $\nu Z_{{\hookrightarrow}q\bar{q}}$ and $eW_{{\hookrightarrow}q\bar{q}}$ channels of the $\nu^*$ search.

\begin{figure}[htbp]
\begin{center}
\includegraphics[width=14pc]{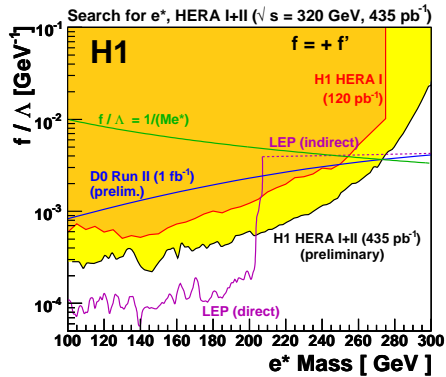}\hspace{2pc}%
\begin{minipage}[b]{14pc}\caption{\label{fig:estar}Exclusion limits on the coupling $f/\Lambda$ at $95$\% C.L. as a function of the mass of the excited electron with the assumption $f = +f'$. 
}
\end{minipage}
\end{center}
\end{figure}

The event yields observed in each decay channel are summarised in table~\ref{tab:exc_fermions}. They are in good agreement with SM expectations and no deviation is observed in invariant mass distributions of excited neutrino candidates.
The new limit on the $e^*$ mass as a function of $f/\Lambda$ is presented in figure~\ref{fig:estar}, using the assumption $f = + f'$. 
Assuming $f/\Lambda = 1/M_{e^*}$ and $f = + f'$, $e^*$ masses below $273$~GeV are ruled out. 

This result is equivalent to the most recent limit on $e^*$ production obtained at the Tevatron by the D0 Collaboration and using $1$~fb$^{-1}$ of data.

%

\end{document}